# A 10-100 GHz Coax to Double-Ridged Waveguide Launcher and Horn Antenna

*Matthew A. Morgan, and Tod A. Boyd*

**Abstract**

The design, fabrication, and measurement of a coax to double-ridged waveguide launcher and horn antenna is presented. The novel launcher design employs two symmetric field probes across the ridge gap to minimize spreading inductance in the transition, and achieves better than 15 dB return loss over a 10:1 bandwidth. The aperture-matched horn uses a half-cosine transition into a linear taper for the outer waveguide dimensions and ridge width, and a power-law scaled gap in order to realize monotonically-varying cutoff frequencies, thus avoiding the appearance of trapped mode resonances. It achieves a nearly constant beamwidth in both E- and H-planes for an overall directivity of about 16.5 dB from 10-100 GHz.

## I. Introduction

Broadband, single-polarization horn antennas are of significant interest in a number of test and measurement applications, and many successful designs based on double-ridged geometry have been demonstrated in the cm-wave frequency band [1]-[6]. Horns up to 18 GHz are especially common and readily available for purchase from commercial vendors. Very few such broadband horns, however, have extended much into the millimeter-wave range. In this paper we present the design, fabrication, and testing of a double-ridged horn with good return loss and near-constant beam-width covering 10-100 GHz. This is believed to be the highest operating frequency for a decade-bandwidth horn reported to date. It is enabled in part by a novel coax to double-ridged waveguide launcher which is described in Section II of this report. The remainder of the horn is described in Section III, and measurements on the complete assembly are presented in Section IV.

## II. Coax to Double-Ridged Waveguide Launcher

The launcher is a critical element in the design of any double-ridged horn, and is key to the horn's overall performance. It must be capable of launching the dominant mode of the double-ridged waveguide with minimal coupling to higher-order modes while presenting a near-constant impedance to the input port over the desired frequency range, in this case a 1mm coaxial connector [7]. This is a challenging design problem in its own right and has received a great deal of attention in the published literature [8]-[10].

First, one must select waveguide dimensions. Ideally, one would like to use a waveguide in the launcher which is single-mode over the entire desired frequency range of operation, however this can lead to very extreme aspect ratios in the gap of the waveguide. A somewhat relaxed condition excludes only those higher-order modes which may be coupled to the dominant one by symmetric perturbations of the waveguide. For this design, we will assume that the horn is symmetric about both E- and H-planes. (It will be shown later that the launcher, strictly speaking, is only symmetric in one-dimension, but the asymmetries lie in the electrically small gap region and are very weakly coupled to the higher-order modes.) The waveguide dimensions are labeled in the inset of Fig. 1. For this design, we have chosen $a/b = 1.7$ and $w/b = 0.7$ in order to obtain the largest possible gap (to ease fabrication) for the given frequency range while satisfying the aforementioned modal characteristics. A plot of the cutoff frequencies, normalized to the dominant mode, is shown in the figure as a function of gap height ratio ($b/g$). The plot shows that a gap height ratio of at least 15 is needed to achieve the desired decade of bandwidth. In practice, we have used $b/g \approx 22$ at the launcher in order to provide a reasonable impedance match down to the lowest operating frequencies. The final waveguide dimensions chosen were $a = 4.318$ mm, $b = 2.54$ mm, $w = 1.778$ mm, and $g = 0.114$ mm.

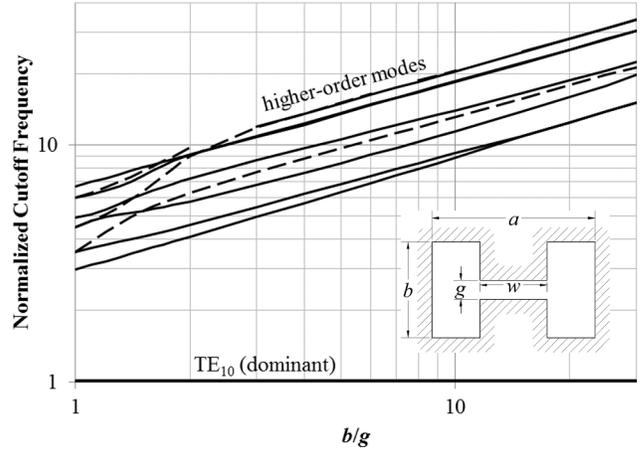

Fig. 1. Mode cutoff frequencies versus gap height ratio ($b/g$) for double-ridged waveguide, normalized to the cutoff of the dominant mode. $a/b = 1.7$ and $w/b = 0.7$. Modes which do not exhibit the dominant electric and magnetic symmetry conditions have been excluded.

The initial design of the launcher comprised a single coaxial probe extending downward from the top ridge, the center-conductor making contact with the bottom ridge across the gap, with a rectangular waveguide backshort. The backshort had the same outer dimensions as the ridged waveguide, but without the ridges, and roughly the same length as the ridges are wide. Unlike a conventional waveguide backshort which behaves like a transmission line stub and would have unacceptable phase rotation over a 10:1 bandwidth, the truncation of the ridges at this junction presents a very high impedance (approximately open-circuit) to the field probe. Attempts to optimize this structure, however, revealed a persistent positive reactance that could not be compensated for. A smith chart of the resultant impedance is shown in Fig. 2a for 10-100 GHz. Note that the impedance curve clusters around a point at constant positive reactance (not constant inductance), suggesting a current path with length inversely proportional to frequency. Close examination of the current density on the surfaces of the waveguide, Fig. 2b, reveals just such a path in the form of the spreading current around the central post of the coaxial port. The opening angle of the fan-out region of the current density is found to be roughly proportional to frequency, hence a frequency-dependent inductance, or constant positive reactance.

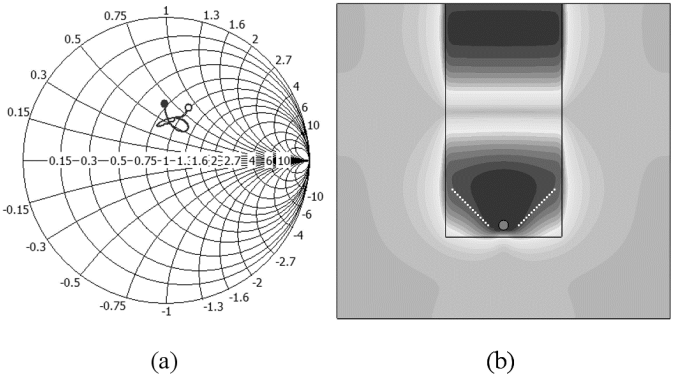

(a) (b)

Fig. 2. Initial optimization of double-ridged launcher. (a) Optimized input impedance as seen from the coaxial port over 10-100 GHz. (b) Snapshot of the current density on the ridge and waveguide walls (top-down view). The white dotted lines highlight the spreading of the current across the width of the ridge from the central post of the coaxial probe. The opening angle of this fan-out region is found to be roughly proportional to frequency.

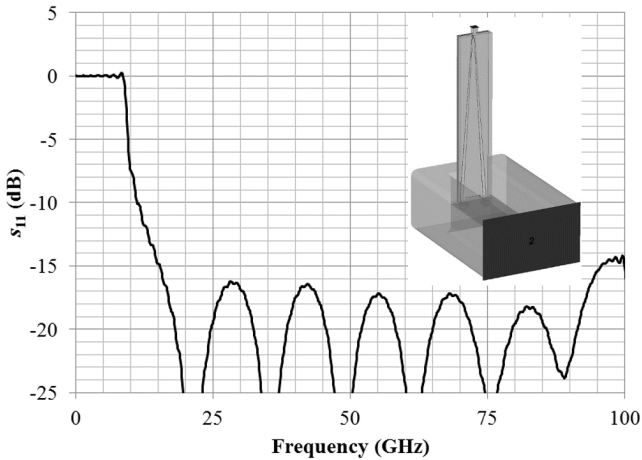

Fig. 3. Model of double-post launcher design and simulated performance.

To combat this spreading inductance, the launcher was modified to use two parallel posts instead of one, thus minimizing the distance over which the current would have to travel from the posts to span the width of the ridge in the waveguide section. Simulations confirmed that this arrangement yields a nearly constant, real-valued impedance.

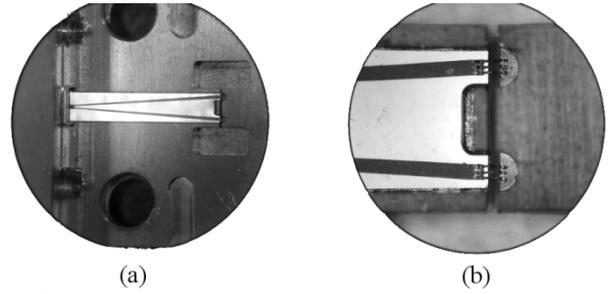

(a) (b)

Fig. 4. Photographs of (a) the launcher assembly with quartz splitter, and (b) bond-wire coupling posts crossing the ridge gap.

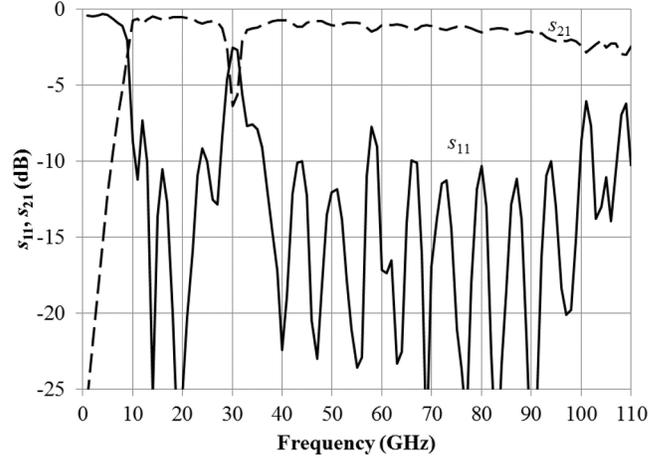

Fig. 5. Measured s-parameters of back-to-back coax to double-ridged waveguide transitions.

This impedance was transformed to 50 Ω via an electrically-long microstrip splitter with linear tapers. The splitter was printed on a 76 μm thick quartz substrate, and the "posts" were replaced by three bond-wires each at the end of the microstrip circuit. An electromagnetic model and simulation of the final launcher assembly is shown in Fig. 3. A photograph of the microstrip substrate and bond-wire coupling posts is shown in Fig. 4.

Two launcher assemblies were fabricated as separate pieces from the horn so that they could be characterized independently in a back-to-back configuration. The very small gap in the double-ridged waveguide cross-section required very precise alignment at the waveguide flanges, which was achieved using precision-ground, "in-plane" alignment pins laid in numerically-machined slots at each junction [11]. Network Analyzer measurements were performed on the back-to-back transitions, shown in Fig. 5. The return loss peaks are expected to be 6 dB higher in the back-to-back configuration, making these results consistent with the predicted single transition return loss of 16 dB shown in Fig. 3b. The resonance at 30 GHz is believed to be due to a trapped higher-order mode in the long double-ridged waveguide between the launchers, and thus is an artifact of the measurement that will not be present in the final horn assembly.

### III. Horn Design

Before settling on a horn design, many different transition and taper profiles were simulated. The design goals were to have good return loss and as near as possible to constant directivity over the 10:1 frequency range. It was found that a smooth transition from the uniform double-ridged waveguide into the main flare of the horn, having no breaks in the slope of the outer walls, was crucial to achieving the best broadband impedance match. A linear primary taper was then selected in order to provide constant directivity. Finally, a circular quarter-turn "rollout" was added to the aperture edges in order to reduce diffraction and provide the best beam characteristics as well as improved return loss [12].

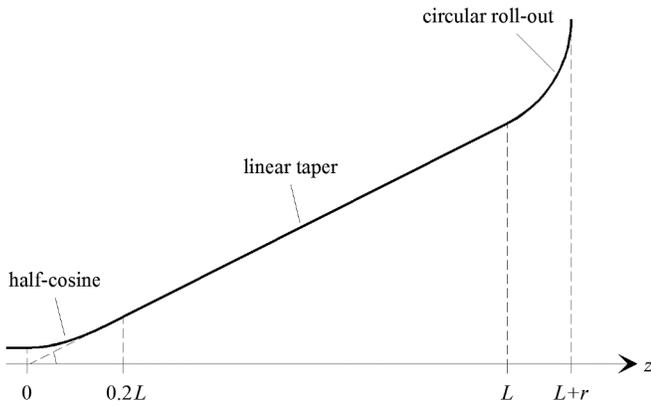

Fig. 6. Diagram of horn taper profile, which applies to the outer dimensions, *a* and *b*, as well as the ridge width, *w*. The gap height was scaled against the height according to a power law in order to ensure a monotonic decrease in cutoff frequencies for all modes.

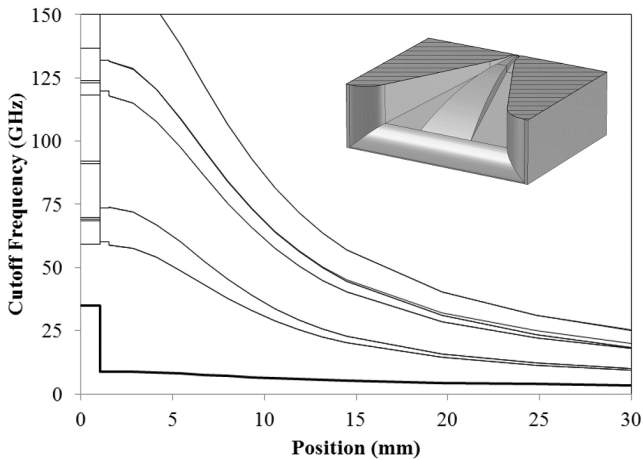

Fig. 7. Trapped mode analysis for the double-ridged horn. The cutoff frequencies for multiple modes are plotted as a function of position along the longitudinal axis of the horn from the launcher at the left to the aperture at the right. A cutaway view of the horn geometry (bottom half) is shown in the upper right.

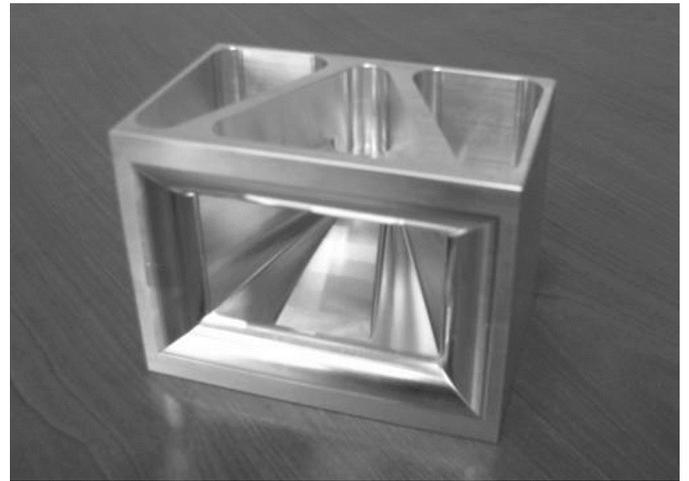

Fig. 8. Photograph of completed horn assembly. Main body measures 10.2 x 7.6 x 6.4 cm.

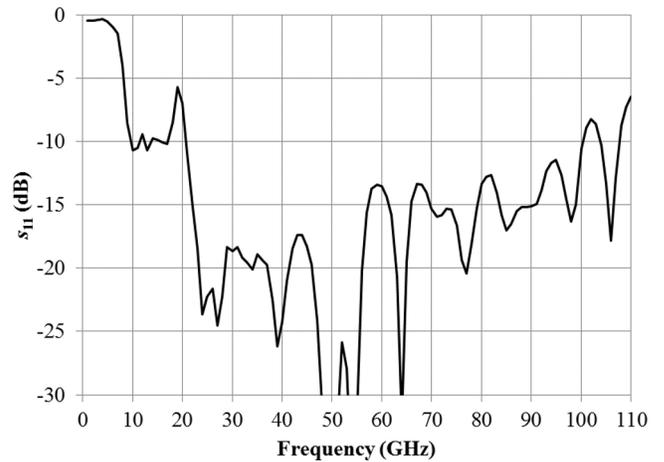

Fig. 9. Measured input return loss of launcher and horn assembly.

A diagram of the final taper profile is shown in Fig. 6. It begins with a short half-cosine-shaped transition into the main taper. The transition-region occupies 20% of the total horn length after the launcher, not including the rollout at the aperture edges. The main portion of the taper profile after this transition is linear. The final aperture dimensions are 15 times larger than that of the input waveguide, having the same aspect ratio, and equally as long as it is wide, or $L \approx 6.5$ cm. Finally, the edges of the aperture were curved away from the taper in circular arcs which maintain the continuity of the slope and terminate in a plane perpendicular to the longitudinal axis of the horn, having added a distance $r = 8.636$ mm to the overall length.

This same basic profile was used, appropriately scaled, for all waveguide dimensions except for the gap. It was seen in Fig. 1 that the cutoff of all higher order modes in the double-ridged waveguide, when normalized to the dominant mode, depend in an approximately power-law fashion on the waveguide height to gap ratio, $b/g$ – that is, the curves appear roughly linear on a log-log plot. Thus, by scaling the gap according to an inverse power-law with respect to the total waveguide height, one can be reasonably assured that if the dominant mode cutoff frequency is monotonic, the same will be true of all modes of higher order. This is crucial to avoiding trapped-mode resonances in a broadband horn. To verify that this would be the case, a trapped mode analysis was performed as described in [13]. The results are plotted in Fig. 7, which shows the expected monotonic cutoff behavior. The absence of local minima in any cutoff frequency curve excludes the possibility of trapped-mode resonances with this structure.

## IV. Horn Measurements

The horn was fabricated as a split block machined from aluminum. The two halves were precisely aligned using the same in-plane dowel pins as the flanges on the coax to double-ridged waveguide launcher. A photograph of the completed horn is shown in Fig. 8. The return loss of the horn and launcher together was measured with a piece of ECCOSORB HR-25 absorber covering the aperture. The results are shown in Fig. 9.

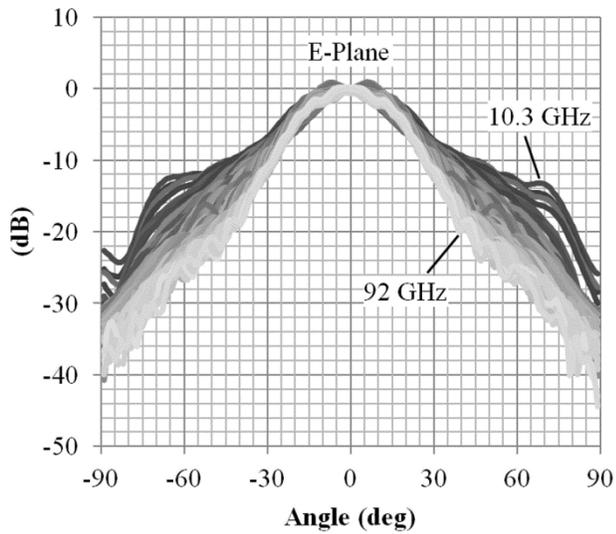

Fig. 10. Measured E-Plane beam patterns for 10-100 GHz double-ridged horn.

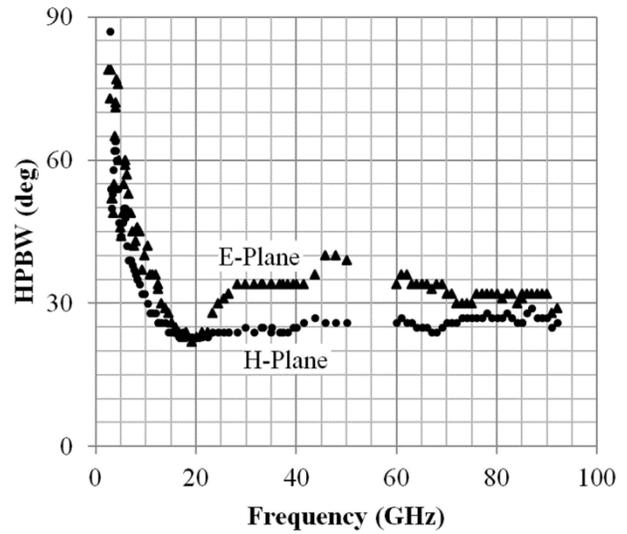

Fig. 12. Measured half-power beam-widths in the E-Plane (triangles) and H-Plane (circles).

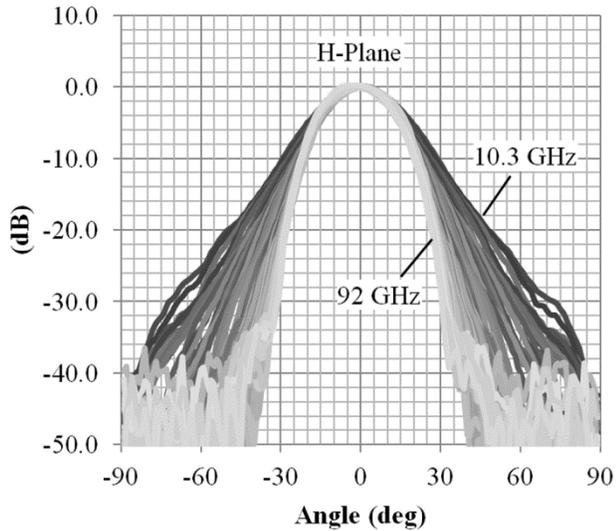

Fig. 11. Measured H-Plane beam patterns for 10-100 GHz double-ridged horn.

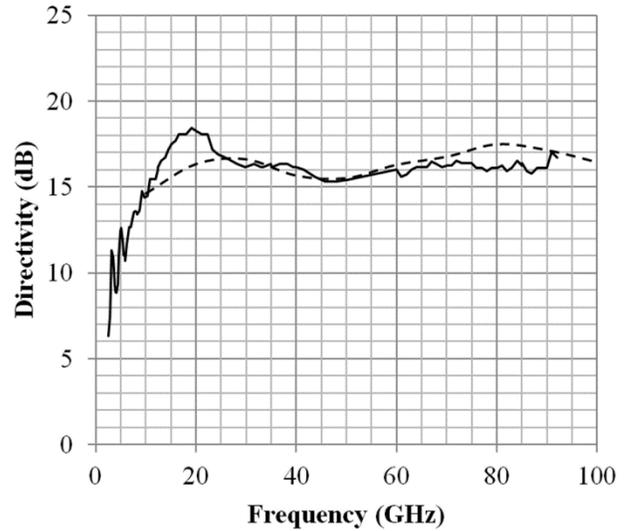

Fig. 13. Modeled (dashed line) and measured (solid line) directivity of 10-100 GHz double-ridged horn.

The beam patterns of the horn were measured in E- and H-Plane cuts in an anechoic chamber. The very broad bandwidth of the horn required patterns to be taken in multiple bands. Eight different standard gain horns were used from 2.6-50 GHz, plus one with WR-12 input for 60-75 GHz, and another WR-10 horn for 75-92 GHz. Beam-pattern measurements were not available between 50-60 GHz or above 92 GHz. The results of these tests are summarized in Fig. 10 and Fig. 11.

Recall that a design criterion for the horn was to have nearly-constant directivity on bore-sight over a 10:1 bandwidth. To calculate the directivity without access to full two-dimensional beam-patterns, the half-power beam widths shown in Fig. 12 were taken from the E- and H-Plane cuts, and then used to estimate the directivity using the formula given in [14],

$$D \approx \frac{DB}{HP_E \cdot HP_H} \quad (1)$$

where the directivity-beamwidth product, $DB$, is assumed to be that of cosine-uniform rectangular aperture (that is, $DB=35,230$ deg$^2$). The measured and modeled directivity is thus shown in Fig. 13. The data is in excellent agreement with theory and shows a constant directivity of about 16.5 dB ± 1.5 dB from 10-100 GHz.

## V. Conclusion

A decade bandwidth double-ridged waveguide launcher and horn have been demonstrated, for the first time in the millimeter-wave range. Such horns can be used in a number of test and measurement applications, including characterization of other antennas, materials, and quasi-optical components. The motivation for this work was to prove a high-frequency broadband horn concept as a scale model for an even higher-frequency design covering 100 GHz to 1 THz, with integrated active electronics as a broadband noise calibration source on the Atacama Large Millimeter and Sub-millimeter Array (ALMA) radio telescope.

## Acknowledgment

The authors wish to thank the National Institute of Standards and Technology (NIST) for calibrated network analyzer measurements in 1mm coax, and their colleague Carla Beaudet for assistance performing beam pattern measurements in the anechoic chamber in Green Bank, WV. The National Radio Astronomy Observatory is a facility of the National Science Foundation (NSF) operated under cooperative agreement by Associated Universities Inc.